# *Gravitational Mass, Its Mechanics – What It Is; How It Operates*

*Roger Ellman*

Abstract

The subject is treated in two parts, inertial mass and gravitational mass. The first part, *Inertial Mass, Its Mechanics – What It Is; How It Operates*, developed inertial mass. The present paper is the second part: *Gravitational Mass, Its Mechanics – What It Is; How It Operates*.

The behavior of gravitation is well known, described by Newton's Law of Gravitation. But what gravitational mass is, how gravitational behavior comes about, what in material reality produces the effects of gravitational mass, is little understood. The extant hypotheses include Einstein's general relativity's bending of space, efforts to develop "quantum gravitation", and attempts to detect "gravitons" and "gravitational waves". Those do not address the mechanics of gravitation but, rather, are abstractions away from it.

From a start of only the limitation on the speed of light, the necessity of conservation, and the impossibility of an infinity in material reality, the present paper presents a comprehensive analysis of the phenomenon gravitational mass:

- how it appears in particles,

- how the Newtonian gravitational behavior arises from that, and

- how the values of inertial mass and gravitational mass are identical,

or, in other words, the mechanics of gravitational mass and gravitation.

Roger Ellman, The-Origin Foundation, Inc.
http://www.The-Origin.org
320 Gemma Circle, Santa Rosa, CA 95404, USA
RogerEllman@The-Origin.org

# *Gravitational Mass, Its Mechanics – What It Is; How It Operates*

*Roger Ellman*

Abstract

## 1. Development of the Unified Field

The following "thought experiments" develop the concept.

Electric Field

- Nothing can travel faster than the speed of light, $c$. Given two static electric charges separated and with the usual Coulomb force between them, if one of the charges is moved the change can produce no effect on the other charge until a time equal to the distance between them divided by $c$ has elapsed.
- For that time delay to happen there must be something flowing from the one charge to the other at speed $c$ and the charge must be the source of that flow.

  The Coulomb effect is radially outward from the charge, therefore every charge must be propagating such a flow radially outward in all directions from itself, which flow must be the "electric field".

Motion of Charge and "At Rest"

- Comparing two such charges, one moving at constant velocity relative to the other, at least one of the charges is moving with some velocity, $v$.
- The flow (of "field") outward from that charge must always travel at $c$. Forward it would go at $[c + v]$ if propagated at $c$ from the source charge already moving that way at $v$. Therefore, it must be sent forward from the charge at $[c - v]$ so that it will travel at $c$ when the $v$ of its source charge is added.

(1) $$Flow_{fwd} = [c - v]$$

- Analogously, rearward it would go at speed $[c - v]$ if propagated at $c$ from the source charge already moving the opposite way at $v$. Therefore, it must be emitted rearward from the charge at $[c + v]$ so that it will travel at speed $c$ when the $v$ of the source charge in the opposite direction is subtracted.

(2) $$Flow_{rwd} = [c + v]$$

- But, that rearward – forward differential means that the direction and speed of motion can be determined by looking at the propagation pattern of the flow as propagated by the charge.

  And, if the pattern were the same in all directions then the charge would be truly "at rest", which means that there is an absolute "at rest" frame of reference.

Unification of Fields

- Except for the kind of field, all of the preceding applies in the same way and with the same conclusions for magnetic field and gravitational field as for electric field.
- Therefore, either a particle that exhibits all three such fields, as for example a proton or an electron, is a source of three separate and distinct such flows, one for each field, or there is only a single flow which produces all three effects: electric, magnetic, and gravitational.

  The only reasonable conclusion is that electric, magnetic, and gravitational field are different effects of the same sole flow from the source particles.

Sources & Their Decay

- The flow is not inconsequential. Rather, it accounts for the forces, actions and energies of our universe.
- For a particle to emit such a flow the particle must be a source of whatever it is that is emitted outward. The particle must have a supply of it.
- The process of emitting the flow from a particle must deplete the supply resource for the particle's emitting further flow, must use up part of its supply, else we would have something-from-nothing and a violation of conservation.

  It must be concluded that an original supply of that which is flowing came into existence at the beginning of the universe and has since been gradually being depleted at each particle by its on-going outward flow.

That Which is Flowing

- The flow is a property of contemporary particles. Those particles are evolved successors to the original oscillations with which the universe began. Then, that which is flowing is the same original primal "medium", the substance of the original oscillations at the beginning of the universe.

  Since it is flowing outward from the myriad particles of the universe simultaneously and that flow is interacting with myriad others of those particles without untoward interference, the "medium" must be extremely intangible for all of that to take place, any one particle's flow flowing largely freely through that of other particles, as intangible as -- well -- "field".

The Beginning

- Before the universe began there was no universe. Immediately afterward there was the initial supply of medium to be propagated by particles. How can one get from the former to the latter while: (1) not involving an infinite rate of change, and (2) maintaining conservation ?

  The only form that can accommodate the change from nothing to something in a smooth transition without an infinite rate of change is the oscillatory form of equation *(3a)*, below.

*(3a)* $U_0 \cdot [1 - \mathrm{Cos}(2\pi \cdot f \cdot t)]$

  The only way that such an oscillation can have come into existence without violating conservation is for there simultaneously to have come into existence a second oscillation, the negative of equation *(3a)* as in equation *(3b)*.

*(3b)* $-U_0 \cdot [1 - \mathrm{Cos}(2\pi \cdot f \cdot t)]$

  That is, the two simultaneous oscillations must have been such as to yield a net of nothing, the prior starting point, when taken together.

The Oscillatory Medium Flow ≡ Electric charge and field

- The initial medium supply of each particle, each being a direct "descendant" of the original oscillation at the universe's beginning, must be oscillatory in form per equations *(3)*. Therefore the radially outward flow from each particle is likewise an oscillatory medium flow of the form of equations *(3)*.

  The flow is radially outward from the particle, therefore, the oscillation of the medium supply of each particle is a spherical oscillation. The particle can also be termed a *center-of-oscillation*, which term will also be used here.

- The amplitude, $U_0$, of the *[1 - Cosine]* form oscillation is the amplitude of the flow emitted from the source particle, which flow corresponds to the electric field. Thus the oscillation amplitude must be the charge magnitude of the source particle -- the fundamental electric charge, $q$, in the case of the fundamental particles, the electron and the proton.

Then, the conservation-maintaining distinction of amplitude $+U_0$ versus amplitude $-U_0$ must be the positive / negative charge distinction.

The frequency, $f$, of the $[1 - Cosine]$ form oscillation must then correspond to the energy and mass of the source particle, that is the energy of the oscillation is $E = h \cdot f$ and the mass is $m = {}^{E}/_{c^2} = {}^{h \cdot f}/_{c^2}$.

[- While it does not pertain to the universe's beginning, because the outward medium flow from each particle must deplete the source particle's remaining supply of medium for further propagation, the amplitude magnitude, $U_0$, must exponentially decay. That is, it must be of the form of equation $(4)$, below.]

$$(4) \qquad |U(t)| = U_0 \cdot \varepsilon^{-t/\tau}$$

Medium Emission and Medium Flow

- When a charge is at rest, medium is emitted by it and flows outward in the same manner in all directions. But, when the charge is in motion at constant velocity, $v$, the flow forward is emitted at speed $[c - v]$ and rearward at $[c + v]$ per above.

- There can be only one frequency, $f$, in the $[1 - Cosine(2\pi \cdot f \cdot t)]$ form oscillation of the emitted flow regardless of whether it is directed forward, rearward or sideward. Therefore, to obtain the slower speed, $[c - v]$, emitted forward the wavelength forward, $\lambda_{fwd}$, must be shorter so that the speed at which the flow is emitted, $= f \cdot \lambda_{fwd}$, will be slower.

- The case is analogous rearward where $\lambda_{rwd}$ is longer in order for the speed, $[c + v]$, to be greater.

In all directions from the moving charge, including any that are partially sideward plus partially forward or rearward, the speed of emission and the wavelength emitted will be the resultant of the sideward plus forward or rearward components of a ray in that direction.

- The absolute rate of flow outward of the emitted medium must be at speed $c$. Forward that comes about because the forward speed of the charge, $v$, adds to the forward speed at which the medium is emitted, $[c - v]$, resulting in the medium flowing at the speed of the sum, $speed = v + [c - v] = c$.

- That speed increase raises the $[1 - Cosine(2\pi \cdot f \cdot t)]$ form oscillation frequency (per the Doppler Effect). Thus forward medium flow speed is $c = f_{fwd} \cdot \lambda_{fwd}$.

- Analogously rearward the speed of medium flow is at $c = f_{rwd} \cdot \lambda_{rwd}$.

In all directions from the moving charge, including any that are partially sideward plus partially forward or rearward, the speed of flow will be $c$ and the frequency and wavelength of the flow will be the resultant of the sideward plus forward or rearward components of a ray in that direction.

Magnetic Field

- A charge at rest exhibits the electrostatic effect but not the magnetic effect. That charge has a spherically uniform pattern of $[1 - Cosine(2\pi \cdot f \cdot t)]$ form oscillatory medium emission and flow outward.

- A charge in motion exhibits the magnetic effect in addition to the electrostatic effect. That charge has a pattern of emission and outward flow of medium that is cylindrically symmetrical about the direction of motion but that varies in wavelength and frequency from $f_{fwd} \cdot \lambda_{fwd}$ forward to $f_{rwd} \cdot \lambda_{rwd}$ rearward.

> The electrostatic [Coulomb's Law] effect is due to charge location. The magnetic [Ampere's Law] effect is due to charge motion. Clearly, then, the electrostatic effect is due to the spherically uniform medium flow from the charge and the magnetic effect is due to the change in shape of that medium flow pattern caused by the charge's motion.

Electro-magnetic Field

- There is a continuous emission of medium in $[1 - \text{Cosine}(2\pi \cdot f \cdot t)]$ oscillatory form from each charge, which medium flows outward, away, forever. Constant velocity motion of a charge produces a change in the frequency and wavelength of that medium flow.

- Changes in the velocity of the charge cause corresponding further changes in the medium's oscillatory form as successive increments of medium are emitted and flow outward from the charge. Earlier increments so changed propagate on outward away from the charge, forever, at $c$. The stream of outward flowing medium carries a history of the motions of the source charge.

> Propagating electromagnetic field is the carrying of both of those field aspects as an imprint on the otherwise uniform medium flow from the charged particle, an imprint analogous to the modulation of a carrier wave in radio communications.

- Electro-magnetic field is caused by acceleration / deceleration of charge, that is by changes in the charge velocity. Therefore:

> The changing electric and magnetic fields of electro-magnetic field actually are form changes imprinted onto the outgoing medium flow and carried passively with it [analogous to modulation of a carrier wave in radio communications].
>
> Because all medium flow is spherically outward in all directions from its source charge, changes in it, caused by changes in the source charge velocity, propagate outward in all directions. Those medium flow changes are the changing electric and changing magnetic fields of electro-magnetic field.

- It is not the speed of light which is the fundamental constant, $c$, light being a mere modulatory imprint on medium flow. It is the speed of medium flow which is the fundamental constant, $c$.

Gravitational Field

- As pointed out earlier above, the frequency, $f$, of the $[1 - \text{Cosine}]$ form oscillation corresponds to the mass of the source particle. Therefore the frequency aspect of the radially outward medium flow is the "gravitational field".

## 2. $\varepsilon$, $\mu$, AND THE SPEED OF PROPAGATION

A brief consideration of an electrical analog to medium wave propagation is necessary to further develop its behavior and the medium. A transmission line is an electrical device for transmitting oscillatory electrical energy from one place to another such as coaxial cables and wire pairs found in radio, and video systems interconnecting components, and the longer length lines to antennas such as wave guides. When electrical signals are introduced at one end of such a line there is a finite speed of travel of the electrical effects along the line.

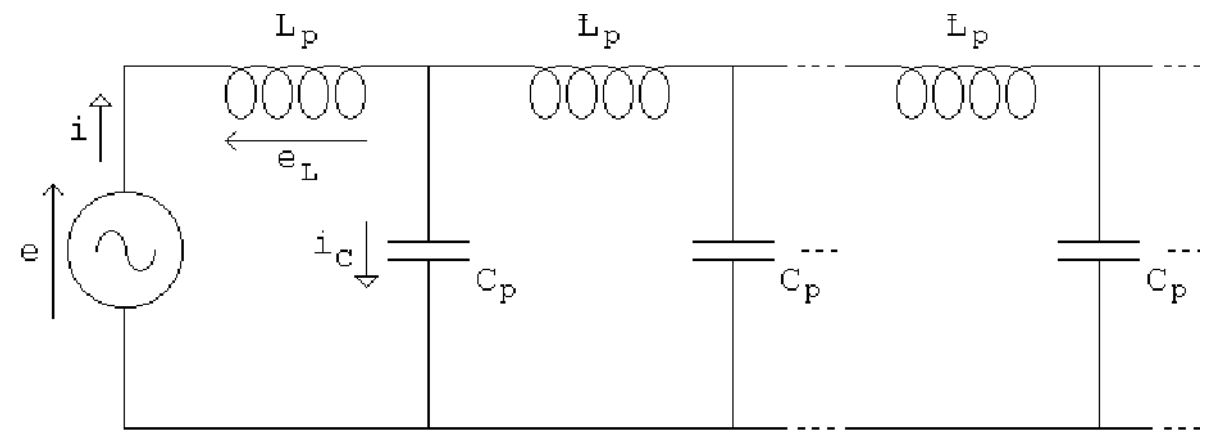

*Figure 1*

The reason is that the line has electrical inductance and capacitance whether placed there or not. These limit the speed of propagation. Since one must deal with various lengths it is best to deal with inductance per unit length of the, $L_p$. Likewise, it is more convenient to deal in terms of the capacitance per unit length of the transmission line, $C_p$.

If an electrical potential source of some magnitude, $e$, is connected to one end of the line, an electric current, $i$, starts to flow in the line. If we consider one quite short segment of the line ("infinitesimally short") it consists (aside from its function as a pure conductor) of an inductance, $L$, and a capacitance, $C$. The amount of the inductance in the segment is

(5) $$L = [\text{Inductance per length}] \times [\text{segment length}] = [L_p] \cdot [v \cdot t]$$

where the length of the short segment, the [segment length] of equation *(5)*, is the product of whatever the speed of travel, $v$, down the line is (its value not yet known) times a minute (infinitesimal) time segment, $t$.

Similarly, the amount of the capacitance is

(6) $$C = [C_p] \cdot [v \cdot t]$$

The potential produced in the inductance is

(7) $$eL = L \cdot \frac{di}{dt} \quad \text{[The natural behavior of inductance, L, in general]}$$

$$= [L_p \cdot v \cdot t] \cdot [{}^{i}/_{t}]$$

$$= L_p \cdot v \cdot i$$

where ${}^{di}/_{dt} = {}^{i}/_{t}$ because $t$ is infinitesimally small.

Similarly, the current through the capacitance is

(8) $$iC = C \cdot \frac{de}{dt} \quad \text{[The natural behavior of capacitance, C, in general]}$$

$$= [C_p \cdot v \cdot t] \cdot [{}^{e}/_{t}]$$

$$= C_p \cdot v \cdot e$$

where, again, ${}^{de}/_{dt} = {}^{e}/_{t}$ because $t$ is so small.

For any such infinitesimal segment of the line, when the electrical potential source, $e$, is first connected to it $e_L = e$, the source voltage, and $i_C = i$, the current in the inductance. That is, all of the voltage appears initially on the inductance and all of the current initially flows into the first segment of capacitance, which initially has no voltage on it. This is most easily visualized for the first instant of time and the first minute segment of the line, segment #1 in the above Figure 1, but, as changes, it is valid for all minute time segments and minute line segments if they are infinitesimal. Therefore

(9) $$e = e_L \quad \text{and} \quad i = i_C$$

$$= L_p \cdot v \cdot i \qquad\qquad = C_p \cdot v \cdot e$$

which, multiplied together, give

(10) $$e \cdot i = [L_p \cdot v \cdot i] \cdot [C_p \cdot v \cdot e]$$

$$= L_p \cdot C_p \cdot v2 \cdot i \cdot e$$

$$v^2 = \frac{1}{L_p \cdot C_p} \quad \text{[Canceling } e \text{ times } i \text{ products and solving for } v^2\text{].}$$

Thus the speed of propagation along the transmission line, the speed of propagation through a medium of distributed inductance and capacitance of values per unit length $L_p$ and $C_p$, is

$$v = \frac{1}{\sqrt{L_p \cdot C_p}} \quad (11) \qquad \text{[Square root of equation 10]}$$

This same result applies to light, which is an electromagnetic propagation through space, which has inductance per unit length of $\mu_0$ and capacitance per unit length of $\varepsilon_0$, as follows.

The electrical inductance of a coil of wire is $\mu_0$ times the dimensions of the coil, $N \cdot A / l$, where $N$ is the number of turns, or loops, in the coil, $A$ is the cross-sectional area of the coil and $l$ is the length of the coil. The $N$ is a dimensionless number. Thus the inductance is, dimensionally, $\mu_0$ times an area divided by a length, that is times a net length. The $\mu_0$ must then be inductance-per-length.

The electrical capacitance of a simple parallel plate capacitor is $\varepsilon_0$, times the dimensions of the capacitor, $A / L$, where $A$ is the area of each of the two identical plates and $L$ is the distance between them. Thus the capacitance is, dimensionally, $\varepsilon_0$ times an area divided by a length, that is times a net length. The $\varepsilon_0$ must then be capacitance-per-length.

Therefore, the speed of light is the already frequently presented

$$v = \frac{1}{\sqrt{\mu_0 \cdot \varepsilon_0}} \quad (12)$$

However, as presented above, the electromagnetic waves of light are merely an imprint, a modulation, on the flowing medium. It is medium that travels at the speed $c$. The speed of light is $c$ only because light, static relative to the flowing medium on which it is imprinted, must travel at the speed of that medium flow. The flowing medium has both its flow and its potential just as does the electrical behavior in the transmission line.

The flowing medium is propagated by the oscillation of the center-of-oscillation. Every oscillation must consist of two aspects so that the oscillation energy is exchanged back and forth between them. Then the propagation of the center involves the propagation of the effect of the two aspects of the center's oscillation. For any oscillation those aspects are always in the form of potential and flow, for example: water pressure "head" and the flow, electrical voltage and current, or a pendulum's potential and kinetic energy.

But where do the $\mu_0$ and $\varepsilon_0$ come from; how does empty "free space", the nothing that was before the universe began, have those characteristics ? It cannot and does not. Until medium appears the "free space" is absolute nothing, the non-existence (except as potentiality to receive medium) of before the origin of the universe. Clearly, it must be the medium itself, the only non-nothing material reality, that is the cause of $\mu_0$ and $\varepsilon_0$.

Propagating medium, then, is like an electrical signal traveling along a transmission line where the electrical signal is the transmission line. Propagating medium waves are medium laying down its own transmission line as it propagates. (It is helpful to picture a humorous animated cartoon analogy of a train going along a railroad track that does not exist in front of the train, imagining a very busy man standing on the front of the locomotive furiously constructing new laid track just in time for the train to roll onto it.)

Three principal quantities are involved in electrical propagation along a transmission line and pertain to each location along the line: the amount of charge at the location, the electric potential and the electric current which is the flow of charge. There are three analogous quantities pertaining to each location in space involved in the propagation of medium: the amount of medium at the location, the medium potential and the medium flow. (This is an analogy and is not to say that medium amount corresponds to charge, and so forth. Rather, it has already been shown that medium flowing corresponds to static charge, that flowing medium produces the static Coulomb effect.)

It is the medium, the amount of medium at a particular location, that carries, and therefore determines, the value of $\mu_0$ and $\varepsilon_0$ at that location. That quantity, the medium amount is a scalar quantity, one having magnitude but not an associated direction, just as is the case with electrical charge. The medium potential, the "head" in analogy to the corresponding water flow term, is a vector quantity, one having both magnitude and direction, and is the impetus, the "driving force" of the medium flow just as electrical potential impels electric charge flow, which is electric current. The medium flow is also a vector quantity, of course.

There is another difference between medium propagation and the transmission line analogy: medium propagates radially outward in all directions from its source whereas a transmission line is linear, does not spread out. The speed of propagation along the transmission line is constant because the inductance and capacitance per unit length are constant. But, since the medium is the source of the $\mu_0$ and $\varepsilon_0$ then as the medium spreads out into ever greater spherical volume the local amount of medium becomes correspondingly diffused so that the values of $\mu_0$ and $\varepsilon_0$ should decrease and the speed of the propagation increase. However, that does not happen, as follows.

The speed of propagation's dependency in a transmission line on the inductance and capacitance really means that the speed of propagation depends on the time that the electrical signal requires to build up its current through the inductance and to build up its voltage on the capacitance, both build-ups being accomplished in each incremental transmission line segment, sequentially segment-by-segment. Since the same identical current flow and charging-up must occur in each successive segment of the transmission line there is no change in the propagation factors, no change in the propagation velocity, from segment to segment.

In the transmission line the electrical input voltage and current are independent of the line's characteristic inductance and capacitance. Consequently a large input signal travels at the same speed in a transmission line as does a small signal. The transmission line, with its independent $L_p$ and $C_p$, literally "just sits there" waiting for whatever input comes along.

But in propagating medium the "input signal", the medium potential and medium flow, are integral with and inseparable from the "line's characteristics", the $\mu_0$ and $\varepsilon_0$ that determine the flow behavior. Medium propagation does not first "lay down its transmission line $\mu_0$ and $\varepsilon_0$ and then propagate medium flow and medium potential into them. Rather the medium propagation is the "laying down" of medium flow-in-$\mu_0$ and medium potential-on-$\varepsilon_0$. Furthermore, the elements of the propagating medium are myriad infinitesimals of flow-in-$\mu_0$ inextricably inter-mixed and inter-existing with myriad infinitesimals of potential-on-$\varepsilon_0$, all distributed in three dimensions. It is a continuum of infinitesimals, not the linear sequence of distributed characteristics that is an electrical transmission line.

For propagating medium the factor determining the speed of propagation is the time required to build up the medium's flow through the $\mu_0$ and to build up its potential on the $\varepsilon_0$. But, in radially outward propagating medium, the flow is inverse square diffused and the potential is inverse square reduced in exactly the same ratio as are the $\mu_0$ and $\varepsilon_0$ reduced. The ratio of that medium wave propagation's medium flow to its $\mu_0$ and of its medium potential to its $\varepsilon_0$ remains constant, and so likewise the speed, radially outward, of its propagation, $c$.

However, that behavior of propagating medium produces a major effect when medium propagating from one source encounters, that is passes through the same space as, medium from some other source. Depending on the orientation of the flows the effect can be a reducing of the speed of propagation, the $c$ of both medium flows.

In general a medium's propagation speed depends on the magnitude of its medium flow (a vector quantity) relative to the amount of its $\mu_0$ (a scalar quantity) and the magnitude of its medium potential (also vector) relative to the amount of its $\varepsilon_0$ (also scalar). Both of the quantities, medium flow and medium potential are proportional to the then, there, local amount of the medium, its amplitude as originally propagated and subsequently inverse-square reduced in its travel to its current location just, as also are the values of $\mu_0$ and $\varepsilon_0$. In speaking of the medium flow to be established through a $\mu_0$ and the medium potential to be developed on an $\varepsilon_0$ one can just as readily speak of the medium amplitude involved in the process of propagating medium through a region of a given $\mu_0$ and $\varepsilon_0$.

If two mutually encountering medium flows are traveling in the same direction their parameters all combine. The combined flow balances the combined $\mu_0$. The combined potential balances the combined $\varepsilon_0$. Just as medium from a single source, diffusing into greater spherical volumes in space maintains constant speed of propagation, $c$, because the ratio of the medium amplitude to the $\mu_0$ and $\varepsilon_0$ remains constant, so two medium flows in the exact same direction have, combined, the same ratio of amplitude to $\mu_0$ and $\varepsilon_0$ as do their individual flows taken separately.

But if those two flows are in exactly opposite directions, through each other, a different result occurs. The $\mu_0$ and $\varepsilon_0$ of the individual flows still combine to produce new somewhat greater values. The $\mu_0$ and $\varepsilon_0$ are scalar quantities, not having direction, and so combine. They simply reflect the local amount of medium regardless of the direction in which it happens to be moving and regardless of the source from which it came.

But, the flows are vectors in opposite directions. They cannot obtain a combining of their medium flows and medium potentials to compensate for the increased values of $\mu_0$ and $\varepsilon_0$ present. They are inverse-square diffusing in opposite directions and there is no valid net resultant. Each flow acts and must act as if alone, as if propagating independently, which it is in its own direction.

This is always the case with vector quantities. Similar vectors can be resolved into a net combination or resultant vector. ("Similar" means that they are identical except for their direction and magnitude, that is they are otherwise the same kinds of quantities, forces for example, and have identical response and cause / effect relations with their environment.) But if the vectors do not linearly respond to their environment in identical fashion such a combination is not valid.

For example electrical signals respond to inductance in proportion to the oscillation frequency of the signal and to capacitance inversely to the frequency. Two electrical signals of different frequencies in an inductive-capacitive environment respond each independently according to its frequency. One cannot take a vector resultant in such a case.

The independent behavior of the components of complex quantities appears in normal human experience. For example, the overall sound of a symphony orchestra is a combination of many waves at many frequencies and would not at all "look" like a simple pure sinusoid if it could be viewed. If it could be viewed it would "look" like the sum, the combination, that it is. But we distinctly and separately hear each of the different instruments, each different note and tone in spite of the sum. Each component, each musical tone and its harmonics acts individually and retains its distinctive character while simultaneously participating in the combination.

Thus, for medium flows propagating in opposite directions through each other there is for each flow an increase in the $\mu_0$ and $\varepsilon_0$ that it must address for which it has available only the same, now proportionally less sufficient, amplitude of its own to drive itself. Each flow encounters a partially reduced value of $c$. Each is effectively slowed by the other.

If the two medium flows encounter each other neither exactly in the same nor opposite directions, they can be resolved into mutually parallel and perpendicular components. The parallel components fit one of the above two cases with the magnitudes being those of the components.

If `u1` is the local medium flow #1 and `u2` the local medium flow #2 then the effect is as in equation *(13)*.

(13) Each of the two flows alone

$$c_1 = f\left[\frac{u_1(\text{amplitude})}{u_1(\mu \text{ and } \varepsilon)}\right] = c$$

$$c_2 = f\left[\frac{u_2(\text{amplitude})}{u_2(\mu \text{ and } \varepsilon)}\right] = c$$

The two flows encountering each other

Same Direction | Opposite Directions

$$c_1 = f\left[\frac{u_1(\text{amplitude})}{u_1(\mu \text{ and } \varepsilon) + u_2(\mu \text{ and } \varepsilon)}\right] < \text{“}c\text{”}$$

$$c_{1,2} = f\left[\frac{u_1(\text{amplitude}) + u_2(\text{amplitude})}{u_1(\mu \text{ and } \varepsilon) + u_2(\mu \text{ and } \varepsilon)}\right] = \text{“}c\text{”}$$

$$c_2 = f\left[\frac{u_2(\text{amplitude})}{u_1(\mu \text{ and } \varepsilon) + u_2(\mu \text{ and } \varepsilon)}\right] < \text{“}c\text{”}$$

Such slowing correlates with gravitational lensing's light path bending. The medium flow carrying light propagation passing a gravitating mass would experience greater slowing of its wave front on the portion nearer to the gravitating mass and lesser slowing further away [because of the inverse square behavior] -- effects tending to bend the direction of the wave front somewhat toward the attracting mass as in the observed gravitational lensing.

## *3. The Mechanism of Gravitation*

This same medium flow slowing behavior that produces the observed bending of light rays by gravitational fields is the fundamental mechanism of gravitation as follows.

Just as the *[1 - Cosine]* type oscillatory wave of propagated medium is the field, so the source of that propagation, the source itself from which the propagation is emitted radially outward in all directions, [hereafter referred to as *center-of-oscillation]* must embody the charge and the mass, that is, the matter of the "particle" involved. Those are the only physical realities underlying our entire universe: the center-of-oscillation and the propagated wave of medium. Those alone are, cause, account for all of matter, mass, field, force, charge, energy, radiation, everything.

A medium-propagating particle, i.e. a center-of-oscillation, being encountered by incoming waves from another such center experiences an unsymmetrical interference with its own normal wave propagation because of those incoming waves. On the side of the encountered center that is toward the source center the encountered center's propagated waves are slowed because they pass through the incoming source waves. On the opposite side of the encountered center there is no such slowing since both the incoming waves (now out-going) and the encountered center's own propagation are in the same direction, away from the source, and at the same speed. See Figure 2, below.

The slowing is the same as that just developed above in conjunction with equation 13. The effect is quite small (as is gravitation) because the inverse square diffusion of the incoming waves makes their amplitude quite small relative to the amplitude of the encountered center's waves immediately adjacent to that center. But the effective result is that the encountered center, forced to propagate at a slightly lesser wave velocity toward the source center, must then move in that direction, its propagation in the opposite direction being forced to accommodate accordingly.

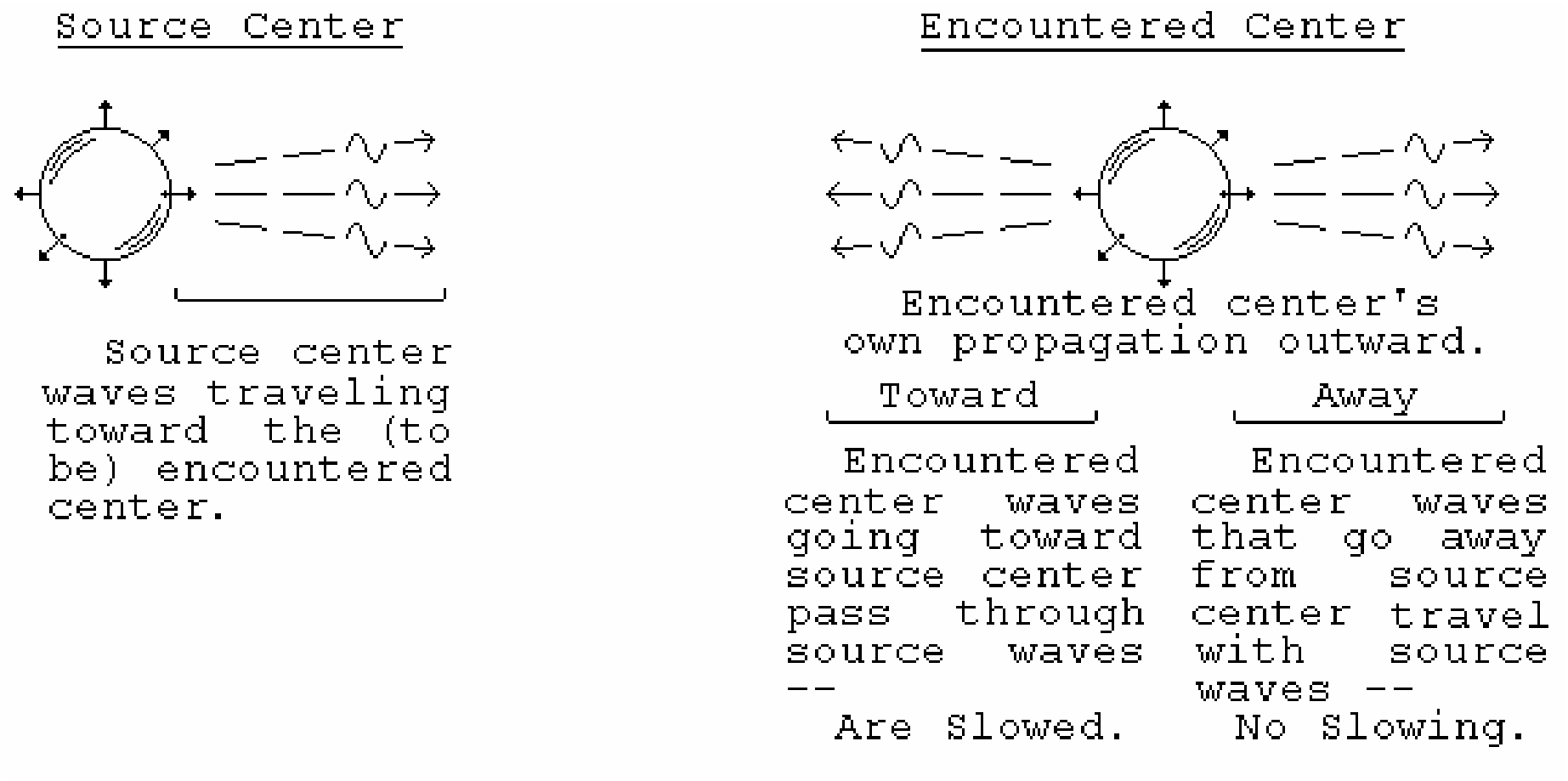

*Figure 2*

That is, a center always propagates forward and rearward as is required by its then state of motion per equations 1 and 2. If an external action, incoming waves, forces a change in that propagation then the state of motion changes, must change, along with the forward and rearward propagation so that all are again consistent with each other.

If the encountered center's wave velocity toward the source center is slowed in the amount $\Delta v$ then the encountered center has been forced by that slowing to perform as follows.

(1) Propagate toward the source at the slowed value,

$$c' = c - \Delta v$$

which requires that

(2) the encountered center must therefore increase its velocity toward the source center to

$$v' = v + \Delta v$$

(where $v$ was the encountered center's velocity toward the source center before the slowing occurred) in order to maintain medium flow at $c$, which further requires that

(3) the encountered center propagate in the direction away from the source (the opposite direction) at

$$c'' = c + \Delta v$$

to compensate.

The incoming waveform producing the slowing is a [1 - Cosine] form as in Figure 3.

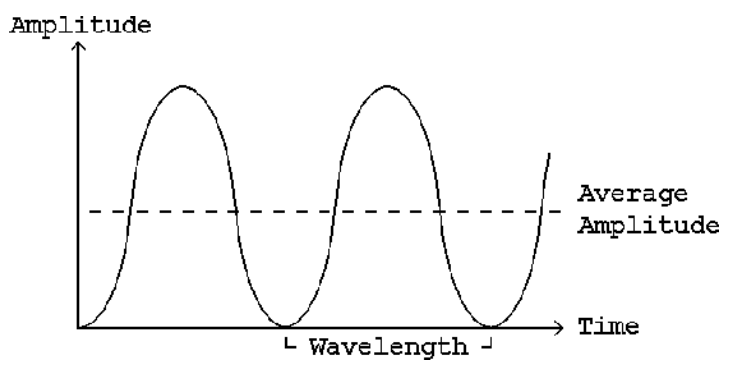

*Figure 3*
*Center Oscillation and Propagated Waveform*

The slowing produced by the incoming waveform arrives as, in effect, pulses. An increment of gravitational slowing of the encountered center's propagation toward the source center, $\Delta v$, occurs for each cycle of incoming wave. Its effect is to change the encountered center velocity by that amount per cycle, each cycle, each period, period-after-period, of the incoming wave's oscillation. That time period is $T_w$, the incoming wave period which is identical to $T_s$, the source center's oscillation period. That is, the frequency, or time rate, of the $\Delta v$ velocity change increments is the source center's oscillation frequency, $f_s$.

As $f_{source}$ is proportional to the source center's mass, $m = h \cdot f / c^2$, $a_{grav}$ is directly proportional to the source center's mass as the gravitational acceleration must be. Because the slowing, $\Delta v$, is directly proportional to the amplitude at the encountered center of the source's waves, $a_{grav}$ exhibits the inverse square reduction as expected and required.

*(a) The Incoming Wave Form from a Proton*

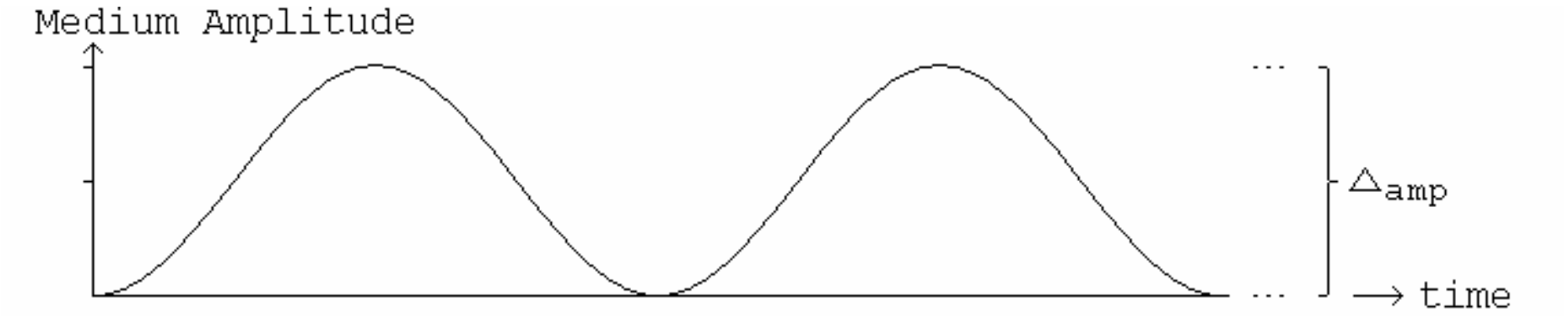

*(b) Its Gravitational Effect on an Encountered Center*

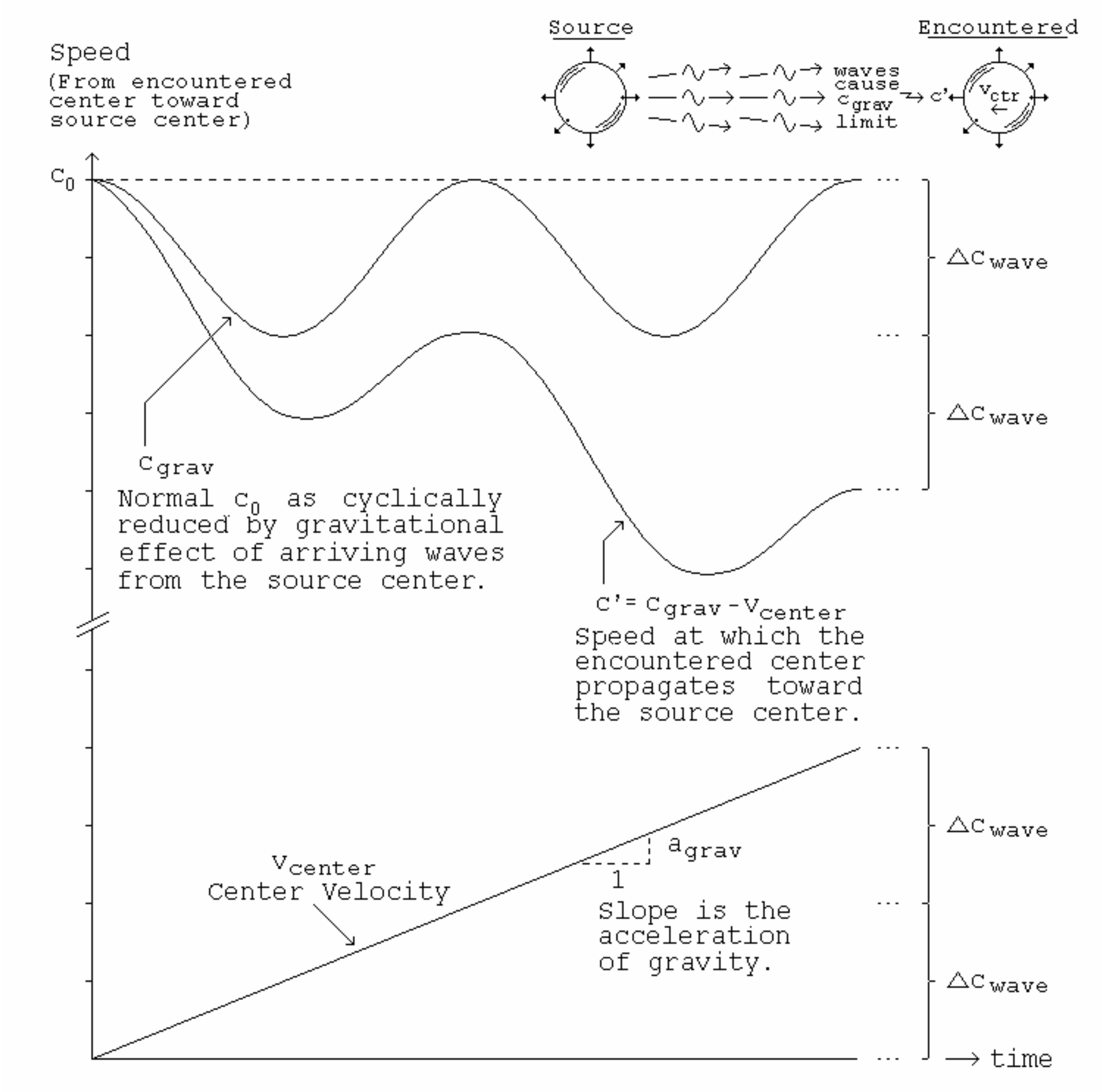

*Figure 4*
*The Gravitational Action*

In the gravitational action a cycle of wave arriving from a distant source center produces a temporary pulse of local increase in the values of μ and ~ on the side of the encountered center that is toward the source center. That reduces the speed at which the encountered center can propagate in the direction of the source center. That imbalance forces the encountered center to make the necessary changes in its velocity and propagation. See Figure 4, above.

The discussion has been in terms of the waves from the source center encountering and producing their effects on the encountered center. Of course all of the effects are mutual. Each of the centers performs in both roles, as source and as encountered center, all of the time and simultaneously. At the same time that the propagation velocity of a center's waves is being reduced by the effect on them of incoming waves – a gravitational action on the center that also results in its acceleration – the outgoing waves of that center are acting similarly on other centers and producing similar effects upon arriving there.

## *4. Analytical Derivation of Newton's Law of Gravitation*

Newton's law of gravitation expressed in terms of $m_{\mathbf{s}ource}$ and $m_{\mathbf{a}cted\text{-}on}$ and with both sides of the equation divided by $m_{\mathbf{a}}$ is, of course,

*(15)*
$$a_{grav} = G \cdot \frac{m_s}{d^2}$$

which states that gravitation is proportional to the mass of the gravitationally attracting body; it is a property of that body's mass.

However, mass and energy are equivalent, so that mass, $m$, is proportional to a frequency, $f$, that is characteristic of that mass. That is

(16)
$$m \cdot c^2 = h \cdot f$$
$$f = \frac{c^2}{h} \cdot m$$

so that the source mass of equation *(1)*, $m_s$, has a corresponding, associated, equivalent frequency, $f_s$.

That being the case, the amount of gravitational acceleration, $a_{grav}$, can be expressed in terms of that frequency as the change, $\Delta v$, in the velocity, $v$, of the attracted mass per time period, $T_s$, of the oscillation at the corresponding frequency, $f_s$, as follows.

*(17)*
$$a_{grav} = {}^{\Delta v}/_{T_s} = \Delta v \cdot f_s$$

It can then be reasoned as follows.

*(18)*
$$a_{grav} = \Delta v \cdot f_s = G \cdot \frac{m_s}{d^2} \qquad \text{[Equating } a_{grav} \text{ of } (15) \text{ and } (17)\text{]}$$

(19)
$$\Delta v \cdot \left[ \frac{m_s}{m_p} \cdot f_p \right] = G \cdot \frac{m_s}{d^2}$$

[Frequency is proportional to mass and $f_p$ and $m_p$ are the proton frequency and mass: $f_s = ({}^{m_s}/m_p) \cdot f_p$.]

$$\Delta v = G \cdot \frac{m_p}{d^2 \cdot f_p}$$

[Rearrange, canceling $m_s$'s.]

Then:

*(20)*
$$\Delta v = G \cdot \frac{1}{d^2 \cdot f_p} \cdot \frac{h \cdot f_p}{c^2}$$

[Substituting $m_p = {}^{h \cdot f_p}/_{c^2}$]

$$\Delta v = G \cdot \frac{h}{d^2 \cdot c^2}$$

Now, the Planck Length, $l_{Pl}$, is defined as

(21) $$l_{Pl} \equiv \left[ \frac{h \cdot G}{2\pi \cdot c^3} \right]^{1/2}$$ [the $^h/_{2\pi}$ part being h-bar]

so that

(22) $$G = \frac{2\pi \cdot c^3 \cdot l_{Pl}^2}{h}$$

Substituting $G$ as a function of the Planck Length from equation (22) into $G$ as in equation (20), the following is obtained.

(23) $$\Delta v = \frac{2\pi \cdot c^3 \cdot l_{Pl}^2}{h} \cdot \frac{h}{d^2 \cdot c^2}$$

$$\Delta v = c \cdot \frac{2\pi \cdot l_{Pl}^2}{d^2}$$ [Simplifying]

This result states that:

- the velocity change due to gravitation, $\Delta v$,
- per cycle of the attracting mass's equivalent frequency, $f_s$,
  - which quantity, $\Delta v \cdot f_s$, is the gravitational acceleration, $a_{grav}$,
- is a specific fraction of the speed of light, $c$, namely the ratio of:
  - $2\pi$ times the Planck Length squared, $2\pi \cdot l_{Pl}^2$, to
  - the squared separation distance of the masses, $d^2$.

That squared ratio is, of course, the usual inverse square behavior.

This result also means that at distance $d = \sqrt{2\pi} \cdot l_{Pl}$ from the center of the source, attracting, mass the acceleration per cycle of that attracting mass's equivalent frequency, $f_s$, namely $\Delta v$, is equal to the full speed of light, $c$, the most that it is possible for it to be. In other words, at that [quite close] distance from the source mass the maximum possible gravitational acceleration occurs. That is the significance, the physical meaning, of $l_{Pl}$ or, rather, of $[2\pi]^{1/2} \cdot l_{Pl}$.

If the original definition of $l_{Pl}$ had been in terms of $h$, not $h\text{-}bar = {}^h/_{2\pi}$ the distinction with regard to $[2\pi]^{1/2}$ would not now be necessary. The $2\pi$ is a gratuitous addition, coming about from the failure to address the Hydrogen atom's stable orbits as defined by the orbital path length being an exact multiple of the orbital matter wavelength. The statement that the orbital electron's angular momentum is quantized, as in

(24) $$m \cdot v \cdot R = n \cdot \frac{h}{2\pi}$$ [n = 1, 2, ...]

is merely a mis-arrangement of

(25) $$2\pi \cdot R = n \cdot \frac{h}{m \cdot v} = n \cdot \lambda_{mw}$$ [n = 1, 2, ...]

the statement that the orbital path, $2\pi \cdot R$, must be an integral number of matter wavelengths, $\lambda_{mw}$, long. And, that may have resulted from a lack of confidence in the fundamental significance of matter waves because of the failure to develop theory that produced acceptable, valid, matter wave frequencies, ones such that $f_{mw} \cdot \lambda_{mw} = \text{particle velocity}$, which is an obvious necessity. That problem is resolved in "A Reconsideration of Matter Waves"[3] where a reinterpretation of Einstein's derivation of relativistic kinetic energy (which produced his famous $E = m \cdot c^2$) leads to a valid matter wave frequency and a new understanding of particle kinetics and of the atom's stable orbits.

The physical significance of $l_{Pl}$ is in its setting of a limit on the minimum separation distance in gravitational interactions and its implying that a "core" of that radius is at the center of fundamental particles having rest mass. That is, equation *(9)* clearly implies that it is not possible for a particle having rest mass to approach another such particle closer than that distance. It is as if that distance is the radius of some impenetrable core of particles having rest mass.

That physical significance of $\sqrt{2\pi \cdot l_{Pl}}$, is so fundamental, fundamental to gravitation and apparently fundamental to particle structure, that it more truly represents a fundamental constant than does $l_{Pl}$. For those reasons that distance should replace $l_{Pl}$ as a fundamental constant of nature as follows.

*(26)* <u>The fundamental distance constant</u> δ.

$$\delta^2 \equiv 2\pi \cdot l_{Pl}^2$$

$$\delta = 4.05084 \times 10^{-35} \text{ meters} \qquad \text{[1986 CODATA Bulletin]}$$

Equation *(9)*, above, then becomes equation *(13)*, below,

*(27)*
$$\Delta v = c \cdot \frac{\delta^2}{d^2}$$

a quite pure, precise and direct statement of the operation of gravitation. It states that gravitation is a function of the speed of light, $c$, and the inverse square law, in the context of the oscillation frequency, $f_s$, corresponding to the attracting, source body's mass. It is interesting to note that equation *(27)* is exact without involving a constant of proportionality such as $G$.

The equation (27) result can also be obtained directly from consideration of solely how slowing is caused by $\mu$ and $\varepsilon$, which demonstrates that the cause of gravitation is the slowing of wave propagation presented in the prior section "*ε, μ, and the Speed of Propagation*" and equation (13). That is as follows.

For propagating medium, at the instant of its propagation from its source center responding to its own $\mu_0$ and $\varepsilon_0$, the value of those two are constant at what we term their free space values. Those values are inverse square reduced as the medium carrying them propagates outward from their source center-of-oscillation. (As discussed in the prior section, the speed of wave propagation remains the same because the waves are also inverse square reduced in amplitude.)

*(28)*
(1) At distance $\delta$ from the center of the source center-of oscillation, the first place where the propagated medium appears and the place where its concentration is greatest, the values of $\mu$ and $\varepsilon$ are the free space values:

$$\mu = \mu_0 \qquad \text{and} \qquad \varepsilon = \varepsilon_0$$

(2) Because of the inverse square law the values at distance *"d"* from the center of the source center-of-oscillation are:

$$\mu(d) = \mu_0 \cdot \frac{\delta^2}{d^2} \qquad \text{and} \qquad \varepsilon(d) = \varepsilon_0 \cdot \frac{\delta^2}{d^2}$$

Then, the overall net effective values when flowing medium from a distant center passes through the outward propagation of an encountered center [per equation *(28),* above] are per equation *(29)*.

(29)
$$\mu_{net} = \left[\mu_0 + \mu_0 \cdot \frac{\delta^2}{d^2}\right] = \mu_0 \cdot \left[1 + \frac{\delta^2}{d^2}\right]$$

$$\varepsilon_{net} = \left[\varepsilon_0 + \varepsilon_0 \cdot \frac{\delta^2}{d^2}\right] = \varepsilon_0 \cdot \left[1 + \frac{\delta^2}{d^2}\right]$$

The resulting net speed of propagation is, then *(30)*

$$c_{net} = \frac{1}{\left[\mu_{net} \cdot \varepsilon_{net}\right]^{1/2}} = \frac{1}{\left[1 + \frac{\delta^2}{d^2}\right] \cdot \left[\mu_0 \cdot \varepsilon_0\right]^{1/2}}$$

$$= \frac{c}{\left[1 + \frac{\delta^2}{d^2}\right]} = \frac{d^2}{d^2 + \delta^2} \cdot c$$

and the amount of the slowing is

(31)
$$\Delta c = c - c_{net}$$

$$= c \cdot \left[1 - \frac{d^2}{d^2 + \delta^2}\right]$$

$$= c \cdot \frac{\delta^2}{d2^2 + \delta^2}$$

$$= c \cdot \frac{\delta^2}{d^2} \qquad [d^2 \text{ is much greater than } \delta^2]$$

so that

(32)
$$\Delta v = c \cdot \frac{\delta^2}{d^2} \qquad [\text{the slowing, } \Delta c\text{, equals the velocity change, } \Delta v]$$

which is identical to equation *(27)*.

With the definition of $\delta$ per equation *(26)* the simplest statement of the universal gravitation constant, $G$, is

(33)
$$G = \frac{c^3 \cdot \delta^2}{h} \qquad [\text{Substituting } \delta \text{ for } 2\pi \cdot l_{Pl}{}^2 \text{ in equation (22)}]$$

The length $\delta$ is a true universal constant joining $c$, $q$, $h$, and $\pi$ as fundamental constants that characterize the universe and completing that family.

Equation *(27)* gives the gravitationally caused velocity change per cycle of the incoming gravitational wave field. The time rate of those velocity change increments, i.e. the gravitational acceleration, $a_g$, is $\Delta v$ times the incoming wave's frequency, which is the source center's frequency, $f_s$, as equation *(34)*.

(34)
$$a_g = \Delta v \cdot f_s$$

$$= c \cdot \frac{\delta^2}{d^2} \cdot f_s$$

$$= c \cdot \frac{\delta^2}{d^2} \cdot \frac{m_s \cdot c^2}{h}$$ [$m_s$ = the source center's mass; $f_s = m_s \cdot c^2/h$.]

$$= G \cdot \frac{m_s}{d^2}$$ [substituting *G* per equation *(33)*]

*(35)* $F_g = a_g \cdot m_e$

$$= G \cdot \frac{m_s . m_s}{d^2}$$ [$m_e$ is the encountered center's mass.]

which is Newton's Law of Gravitation.

As with Coulomb's Law and inertial mass, as developed in the paper, *Inertial Mass, Its Mechanics – What It Is; How It Operates* [2]; here gravitation, gravitational mass, and Newton's Law of gravitation cease to be mere empirically valid observations becoming instead requisite behavior aspects of natural reality derived from fundamentals.

From equation *(34)* it is clear that only $m_s$ operates in the process of gravitation. That is, $m_s$ in equation *(34)* is a gravitational mass and the gravitational acceleration is independent of $m_e$. Equation *(35)* on the other hand is merely a statement of Newton's 2nd Law. The new [relative to equation *(34)*] mass in equation *(35)*, $m_e$, is the inertial mass of Newton's law.

But, the overall action is mutual. The "encountered" center gravitationally attracts the "source" center at the same time as the "source" center gravitationally attracts the "encountered" center. Therefore, the quantities, $m_s$ and $m_e$, are, and operate simultaneously, in both roles – "source" and "encountered", as both types of mass – inertial and gravitational. That means that inertial mass and gravitational mass are identical as follows.

Given two gravitationally attracting bodies, *#1* and *#2,* the force with which *#1* attracts *#2* must equal that with which *#2* attracts *#1* [Newton's 3rd Law of Motion]. That is

*(36)* Using: "Grav" = the gravitation constant [normally "G"]
Body #1 has inertial mass = i
and gravitational mass = g
Body #2 has inertial mass = I
and gravitational mass = G

Then:

$$F_{g,\#1 \Rightarrow \#2} = F_{g,\#2 \Rightarrow \#1}$$

$$\text{"Grav"} \cdot \frac{g \cdot I}{d^2} = \text{"Grav"} \cdot \frac{G \cdot i}{d^2}$$

$$g \cdot I = G \cdot i = k$$ [a temporary constant]

so that:

*(37)* $g = k \cdot i$ and $G = k \cdot I$

Equation *(36)* requires that inertial mass and gravitational mass, if not identical in value, must at least always be in the same ratio, $k$, to each other. Substituting equation *(37)* back into equation *(36)* for *g and G* equation *(38)* results.

*(38)* $F_{g,\#1 \Rightarrow \#2} = F_{g,\#2 \Rightarrow \#1}$

$$\text{"Grav"} \cdot \frac{g \cdot I}{d^2} = \text{"Grav"} \cdot \frac{G \cdot i}{d^2}$$

$$\text{"Grav"} \cdot \frac{[k \cdot i] \cdot I}{d^2} = \text{"Grav"} \cdot \frac{[k \cdot I] \cdot i}{d^2}$$

Therefore, it appears that whether the temporary constant, $k$, equals unity or not is moot because if $k \neq 1$ then whatever value it has is absorbed into the universal gravitational constant "grav" above and normally simply $G$. In other words: inertial mass and gravitational mass are identical.

Further proof of the same is that in the earlier paper, *Inertial Mass, Its Mechanics – What It Is; How It Operates* [2] the inertial mass as used in that paper's inertial derivation is [equation *(17)* of that other paper]

$$m_i = \frac{h/c}{\lambda} = \frac{h \cdot f}{c^2}$$

*[m of equation 17 is inertial mass, $m_i$, from the context]*

and in the present paper gravitational mass is [from the third line of equation *(34)*, above] as a necessary part of the derivation

$$f_s = \frac{m_s \cdot c^2}{h} \quad \text{or} \quad m_g = \frac{h \cdot f}{c^2}$$

*[$m_s$ of equation 34 is gravitational mass, $m_g$, from the context]*

whereby proving that:

*The inertial mass and the gravitational mass are identical.*

[That conclusion, has long been thought by modern physicists to be the case, and has been indicated by the most sophisticated measurements, but has been beyond proof because of the lack of understanding of gravitation.]

## 5. Conclusion

The physical effects that we refer to as gravitational mass and gravitation and their behavior in Newton's Law of Gravitation are all properly interpreted in terms of particles' behavior as propagators of medium waves and the effect of incoming such waves on other such particles and their propagation, slowing their propagation toward the source of the incoming waves.

Gravitational mass is identical to inertial mass.